\documentclass[prl,aps,superscriptaddress,fleqn,showkeys,twocolumn]{revtex4}
\usepackage{amsmath,amssymb,graphicx}
\makeatletter
\renewcommand{\@biblabel}[1]{\hspace*{5mm}}%
\let\oldcite\cite%
\let\oldcitet\citet%
\renewcommand\NAT@split[4]{%
  \gdef\NAT@num{#2}\gdef\NAT@name{#3}\gdef\NAT@date{#2}%
  \gdef\NAT@all@names{#4}%
  \ifx\NAT@noname\NAT@all@names \gdef\NAT@all@names{#3}\fi}
\renewcommand\cite[1]{%
\renewcommand\NAT@open{}\renewcommand\NAT@close{}%
\renewcommand\NAT@sep{;}\NAT@parfalse%
\NAT@numbersfalse%
(\oldcite{#1})}
\renewcommand\citet[1]{%
\renewcommand\NAT@open{(}\renewcommand\NAT@close{)}%
\renewcommand\NAT@sep{,}\NAT@parfalse%
\NAT@numbersfalse%
\oldcitet{#1}}
\makeatother
\begin{document}

\title{Statistical properties of neutral evolution}

\author{Ugo~Bastolla}
\affiliation{Centro~de~Astrobiolog{\'\i}a~(INTA-CSIC),
             28850~Torrejon~de~Ardoz, Spain}

\author{Markus~Porto}
\affiliation{Max-Planck-Institut~f\"ur~Physik~komplexer~Systeme,
             N\"othnitzer~Stra{\ss}e~38, 01187~Dresden, Germany}

\author{H.~Eduardo~Roman}
\affiliation{Dipartimento~di~Fisica and INFN, Universit\`a~di~Milano,
             Via~Celoria~16, 20133~Milano, Italy}

\author{Michele~Vendruscolo}
\affiliation{Department~of~Chemistry, University~of~Cambridge,
             Lensfield~Road, Cambridge CB2~1EW, UK}

\date{July 31, 2002}

\begin{abstract}
Neutral evolution is the simplest model of molecular evolution and thus it is
most amenable to a comprehensive theoretical investigation. In this paper, we
characterize the statistical properties of neutral evolution of proteins under
the requirement that the native state remains thermodynamically stable, and
compare them to the ones of Kimura's model of neutral evolution. Our study is
based on the Structurally Constrained Neutral (SCN) model which we recently
proposed. We show that, in the SCN model, the substitution rate decreases as
longer time intervals are considered, and fluctuates strongly from one branch
of the evolutionary tree to another, leading to a non-Poissonian statistics for
the substitution process. Such strong fluctuations are also due to the fact
that neutral substitution rates for individual residues are strongly correlated
for most residue pairs. Interestingly, structurally conserved residues,
characterized by a much below average substitution rate, are also much less
correlated to other residues and evolve in a much more regular way. Our results
could improve methods aimed at distinguishing between neutral and adaptive
substitutions as well as methods for computing the expected number of
substitutions occurred since the divergence of two protein sequences.
\end{abstract}


\maketitle

\onecolumngrid
\newpage
\twocolumngrid

\section{Introduction}

The molecular clock hypothesis, proposed by Zuckerkandl and Pauling about 40
years ago \cite{Zuckerkandl62}, has been a cornerstone in the foundations of
molecular evolution. In comparing the sequences of homologous proteins,
\citet{Zuckerkandl62} observed that the rate of amino acid substitutions per
site per year in a given protein family is approximately constant,
independently of the pair of species compared. In particular, the substitution
rate appeared to depend only weakly on the number of individuals in the
population and on ecological variables, a property which is extremely useful
for reconstructing phylogenetic trees from the comparison of protein sequences
\cite{Saitou87,Thompson94,Li00}. The constancy of the substitution rate was
shown some years later to be an immediate consequence of the neutral theory of
molecular evolution, proposed independently by \citet{Kimura68} and by
\citet{King69}.

The proposal of the neutral theory raised heated controversies, in part because
it challenged the expectation that most variations in the genetic material are
driven by positive natural selection. At the present time, it is however
generally recognized that both neutral and adaptive substitutions play an
important role in the evolution of protein sequences, and methods are being
developed to distinguish between them. Moreover, the neutral paradigm has been
extended to include slightly deleterious mutations, as proposed by
\citet{Ohta73,Ohta76}. Models of slightly deleterious mutations are however
rather complicate from the point of view of population genetics, and will not
be considered here.

Despite this debate, less attention has been drawn to the question whether
Kimura's neutral model adequately describes the statistical properties of
neutral evolution of proteins. This question is an important one, since an
accurate knowledge of such expected properties would allow to better
distinguish between neutral and adaptive substitutions and to improve methods
to reconstruct evolutionary trees from sequence alignments.

Kimura's model assumes that the overwhelming majority of mutations are either
effectively neutral (their effect on the fitness is much smaller than the
inverse of the effective population size) or lethal, in the latter case purged
by negative selection. We will call this assumption the neutral hypothesis.
Moreover, the rate of occurrence of neutral mutations is regarded as constant
throughout evolution, independent of the current sequence. This is the
homogeneity hypothesis. As a consequence, the rate of neutral substitutions is
predicted to be constant, and the number of neutral substitutions being fixed
after $k$ generations of evolution is predicted to be a Poissonian variable of
mean value $\mu x k$, where $\mu x$ is the neutral mutation rate. At variance
with this, the first tests performed by \citet{Ohta71} on homologous proteins
showed that the variance of the substitution distribution appears to be larger
than that expected for a Poissonian distribution. However, since deviations
seemed at first to be small, the Poissonian statistics was still regarded as a
valid first approximation. More accurate later studies showed that such
deviations are in fact not small \cite{Langley73,Gillespie89}. This discrepancy
between the model and empirical observations was taken by Gillespie as evidence
against the neutral hypothesis, and he favored the hypothesis that most
substitutions in protein sequences are fixed by positive selection
\cite{Gillespie91}. On the other hand, \citet{Takahata87} proposed a
modification of the neutral model, the fluctuating neutral space model, that
can account for the non-Poissonian statistics of substitutions, still
preserving the simplicity of the neutral model as the simplest population
genetics model. This is one of the few instances in which the debate, instead
of concentrating on the neutral hypothesis, reconsidered the way in which
Kimura had modeled neutral mutations.

Recently, we have introduced a model in which the neutrality of mutations in
protein sequences is explicitly tested by means of a computational model of
protein folding \cite{Bastolla99,Bastolla00a,Bastolla02a,Bastolla02b}. The
statistical properties of the model are rather robust: We obtained the same
qualitative behavior by using two folding models as different as a well
designed lattice model \cite{Abkevich94} and a model which allows recognition
of native protein folds \cite{Bastolla00b,Bastolla01}. In particular, we found
that the hypothesis of homogeneity should be rejected and that the neutral
mutation rates fluctuate broadly during evolution. As a consequence, the
variance of the substitution process is much larger than that expected for a
Poissonian distribution, in reasonable agreement with the statistics observed
for most protein families.

Our model belongs to the ``structural approach'' to molecular evolution, which
has been made possible by the recent advances in the understanding of the
dynamics of protein folding and the thermodynamics of biomolecules. In
particular, it is now possible to assess approximately the thermodynamic
stability of biomolecules by computational methods. We have undertaken the
challenge of exploring the applications of these new methods to the field of
molecular evolution, where they can complement more traditional methods based
on population genetics and on sequence comparison. The structural approach has
been pioneered by Schuster and co-workers with a series of studies of neutral
networks of RNA secondary structures \cite{Schuster94,Huynen96,Fontana98} and
it has been applied to proteins by several other groups
\cite{Shakhnovich96,Bornberg-Bauer97,Babajide97,Govindarajan97,Govindarajan98,%
Bussemaker97,Tiana98,Mirny99,Bornberg-Bauer99,Dokholyan01,Xia02}.

In this paper, we study the statistical properties of neutral evolution of
proteins as they are obtained from the Structurally Constrained Neutral (SCN)
model that we have recently introduced \cite{Bastolla02a}. Our goal is to
explore the consequences of these statistical properties on molecular
evolution. We start the paper by reviewing Kimura's model of neutral evolution,
making explicit the homogeneity hypothesis and the arguments that seem to
support its validity. In the next section, we review the SCN model, on which
our results are based. Conservation of the protein fold during evolution is
discussed in the following section, where we argue that fold conservation can
be regarded as a consequence of the requirement of thermodynamic stability. We
then investigate the origin of the strong fluctuations in the fraction of
neutral neighbors that are found within the SCN model and that produce the
non-Poissonian statistics of the substitution process. We define the neutral
connectivities at each position in the protein and study their pairwise
correlations. We find that even connectivities at positions far apart in the
structure are strongly correlated. Remarkably, the only exceptions to this
behavior are represented by structurally conserved positions, which are much
less correlated with other positions and evolve in a more regular fashion.
These spatial correlations are the counterpart of the temporal correlations
that we described in an earlier work \cite{Bastolla02b}. In the following
sections, we present some consequences of the broad distribution of neutral
rates. We show that: (i)~The neutral mutation rate is not constant as a
function of the time interval considered, but decreases monotonously as larger
time intervals are considered. (ii)~The normalized variance of the substitution
process increases with the time interval, and is in reasonable agreement with
the data on the dispersion index for most proteins. (iii)~Fluctuations in the
neutral connectivities can influence the generation time effect. Our
simulations are then used to test methods applied to estimate the number of
substitutions in the divergence of two sequences. Finally, we discuss the
impact of our findings on methods used to detect positive selection (as for
instance the one proposed by \citet{McDonald91}), which are usually based on
neutral models with constant neutral mutation rate. After summarizing our
results in the Conclusions, the section `Materials and Methods' reports the
details of some technical points.

\section{Kimura's neutral model revisited}

As discussed in the Introduction, Kimura's neutral model is based on two
assumptions. The neutral hypothesis is equivalent to the assumption that the
most common mutations in protein sequences are either disruptive and eliminated
by negative selection, or neutral, i.e.\ they leave the protein active and
their effect on fitness is much smaller than the inverse of the effective
population size. This mutational spectrum implies that protein sequences evolve
on a neutral network, i.e.\ a set of sequences where the protein is active and
which can be connected through point mutations. Fixation of slightly
deleterious mutations, as well as advantageous mutations, are not included in
the model. This is of course an important limitation of neutral models.

The second assumption is that the rate of appearance of neutral mutations is
constant throughout evolution (homogeneity hypothesis). In a 1977 paper, Kimura
commented that rate constancy may not hold exactly \cite{Kimura77}, but he did
not develop the consequences of the violation of this hypothesis any further.

The neutral model predicts that the rate of fixation of neutral mutations in an
evolving population is equal to the rate of their appearance, independently of
the population size, because the number of appearing mutations is proportional
to the population size and the probability of their fixation is inversely
proportional to it. Let us consider a given lineage evolving for a time
interval $t$. The number of mutations appearing in this time interval is
expected to obey a Poissonian statistics with mean value $\mu t$, where $\mu$
is the mutation rate. We call \textit{neutral connectivity} the probability
that one of such mutations is neutral, which is proportional to the
connectivity of the neutral network, and denote it by $x$. The value of $x$
depends on the particular protein family chosen but it is assumed to be
constant throughout evolution. As a consequence, the probability
$\mathrm{P}(S_t=n)$ of the occurrence of $n$ neutral mutations within a time
interval $t$ is
\begin{equation}
\mathrm{P}(S_t=n) = \sum_{m=n}^\infty \mathrm{e}^{-\mu t} \frac{(\mu t)^m}{m!}
\binom{m}{n} x^n (1-x)^{m-n} \, ,
\end{equation}
where the factor $\binom{m}{n} x^m (1-x)^{m-n}$ gives the probability that $n$
out of $m$ mutations are neutral. The summation can be performed exactly,
yielding
\begin{equation}
\mathrm{P}(S_t=n) = \mathrm{e}^{-\mu x t} \frac{(\mu x t)^n}{n!} \, ,
\end{equation}
so that the number of neutral mutations also obeys a Poissonian statistics,
with mean value $\mu x t$ and rate $\mu x$.

In principle the neutral connectivity $x(\mathbf{A})$ depends on the amino acid
sequence $\mathbf{A}$ considered, but Kimura's model assumes that
$x(\mathbf{A})$ is the same for all protein sequences belonging to the same
structural and functional family. This assumption can be justified with the
following argument: Let $x_i(\mathbf{A})$ be the fraction of possible mutations
of sequence $\mathbf{A}$ at position $i$ which are neutral. Then the overall
fraction of neutral mutations is just the average of this quantity over the $N$
positions in the amino acid chain,
\begin{equation}
x(\mathbf{A}) = \frac{1}{N} \sum_i x_i(\mathbf{A}) \, .
\end{equation}
If the neutral connectivities at different positions are uncorrelated, or if
only positions which are close in the native structure are correlated, then the
variance of $x(\mathbf{A})$ is proportional to $1/N$, which is very small for
long protein chains, so that the approximation of constant $x$ is expected to
be reasonably good. As we shall see in the following, this is not the case.

\section{The Structurally Constrained Neutral Model}

In the SCN model we assume the validity of the neutral hypothesis, but we do
not make any assumption regarding the neutral mutation rate. Instead, we
estimate explicitly the effect of a mutation on protein stability using an
effective model of protein folding \cite{Bastolla00b} which provides us with a
genotype to phenotype mapping. In this respect, the rate of occurrence of
neutral mutations is an outcome of the model. We show that this rate displays
very broad fluctuations throughout evolution.

The SCN model addresses protein evolution at the level of a single sequence,
and it does not take into account population dynamics. This simplification is
justified by the fact that, within Kimura's model, the substitution rate does
not depend on the population size. Nevertheless, population size might
influence the evolutionary process if the rate of neutral mutations shows broad
fluctuations, as observed here. The explicit inclusion of population genetics
into the model will be needed to investigate this interesting possibility.

\subsection{Estimating protein stability}

In our model of protein folding, we evaluate effective conformational energies
(temperature and pH dependent) using the effective energy function described by
\citet{Bastolla00b,Bastolla01}, which provides good thermodynamic properties
for protein structures in the Protein Data Bank (PDB). For each sequence
typically several millions alternative conformations of the same length are
generated by gapless threading. The native conformation is identified as the
lowest energy conformation. The energy function has several non-trivial
properties: (i)~It assigns lowest energy to the experimentally known native
structures of basically all single chain protein sequences; (ii)~It provides
the native structure with a well correlated energy landscape (see below).
(iii)~For chains belonging to oligomeric proteins, for which interchain
interactions are neglected, the experimentally known native state shows a
deficit of stabilizing energy, well correlated with the amount of neglected
free energy. (iv)~The effective energy function that we use is able to estimate
the folding free energy of proteins with two states kinetics reasonably well
(U.~Bastolla, unpublished).

There are two necessary conditions for thermodynamic stability (a)~a low
folding free energy, i.e.\ the unfolded state is less stable than the native
state; (b)~stability with respect to alternative misfolded states, i.e.\ the
protein has a well correlated energy landscape, a property which turns out to
be crucial in all simple models of protein folding. To enforce the first kind
of stability, we estimate the folding free energy, which, in the simplest
approximation, is just a linear function of the effective energy of the native
state. The second condition is estimated through two computational parameters:
The $Z$-score \cite{Bowie91,Goldstein92}, which measures the difference between
the native energy and the average energy in units of standard deviation of the
energy, and the normalized energy gap $\alpha$ \cite{Gutin95,Bastolla99},
measuring the minimal value of the energy gap between an alternative
conformation and the target one divided by their structural distance. A
positive and large value of the $\alpha$ parameter ensures both that the target
conformation has lowest energy and that the energy landscape is well
correlated, in the sense that conformations very different from the native one
have very high energy.

We impose stability conservation by requiring that the lowest energy state of
the model coincides with the experimentally known native state and allowing its
stability parameters, previously described, to be off not more than 1.5\% with
respect to the values of the corresponding PDB sequence.

\subsection{Exploring the neutral network}

Simulations of protein evolution are performed starting from a protein sequence
in the PDB. Evolution is constrained to viable sequences, which are sequences
where the native structure is thermodynamically stable. A neutral network is a
set of viable sequences which can be connected to the starting sequence through
point mutations passing on other viable sequences. Thus sequences on a neutral
network share the same protein fold and are evolutionarily connected.

For every amino acid sequence $\mathbf{A}$ in the neutral network we measure
the fraction of neutral neighbors $x(\mathbf{A}) \in (0,1]$, which is the
fraction of possible point mutations that are viable. Since this number defines
the connectivity of the neutral network at point $\mathbf{A}$, we shall also
call $x(\mathbf{A})$ the neutral connectivity of sequence $\mathbf{A}$. In the
framework of our model, this is the only property of a sequence which
influences its evolution.

Subsequently visited sequences belonging to the neutral network constitute an
evolutionary trajectory $\{ \mathbf{A}_1, \mathbf{A}_2, \ldots \}$. To each
trajectory is associated the list of the corresponding neutral connectivities
$\{ x(\mathbf{A}_1), x(\mathbf{A}_2), \ldots \}$.

\subsection{Substitution process}

An amino acid substitution is controlled by two independent events: A random
mutation, described as a Poissonian process as in the previous section, and an
acceptance process which consists in testing whether the sequence is viable.
The acceptance probability for a mutation taking place when the protein is in
sequence $\mathbf{A}$ is given by $x(\mathbf{A})$. As a result of the
fluctuations in the neutral connectivities, the resulting substitution process
is no longer Poissonian. For a given sequence of neutral connectivities $\{
x(\mathbf{A}_1), x(\mathbf{A}_2), \ldots \}$ we can compute the probability
that the number $S_t$ of accepted mutations in a time interval $t$ is equal to
$n$. This is just the product of the probability (Poissonian) that $k$
mutations take place in the time interval $t$ times the probability that $n$ of
these are accepted,
\begin{equation}
\mathrm{P}(S_t=n) = \sum_{k=n}^\infty \mathrm{e}^{-\mu t}
\frac{(\mu t)^k}{k!} P_{\mathrm{acc}}(n|k) \, ,
\end{equation}
where the acceptance probability of $n$ mutations out of $k$ is given by
\begin{equation}\label{eq:Pacc}
P_{\mathrm{acc}}(n|k)=
\left( \prod_{i=1}^n x(\mathbf{A}_i) \right)
\sum_{\{ m_j \}} \prod_{j=1}^{n+1} \left[ 1-x(\mathbf{A}_j) \right]^{m_j} \, .
\end{equation}
Here, the $\{ m_j \}$ are all integer numbers between zero and $k-n$ satisfying
$\sum_{j=1}^{n+1} m_j = k-n$. In other words, the probability that a mutation
is accepted is $x(\mathbf{A}_1)$ as long as the protein sequence is
$\mathbf{A}_1$, $x(\mathbf{A}_2)$ as long as the sequence is $\mathbf{A}_2$,
and so on.

Two kinds of random variables have to be distinguished. The first kind of
variables represent the mutation and acceptance process. The average over these
variables is indicated by angular brackets $\big< \cdot \big>$, and it is still
a random variable dependent on the realization of the evolutionary trajectory.
The average over evolutionary trajectories is indicated by an overline
$\overline{\vphantom{S}\cdot\vphantom{S}}$. In the biological metaphor, the
average over mutation and acceptance correspond to population averages and the
average over different evolutionary trajectories correspond to averages over
independent populations.

If all sequences have the same fraction of neutral neighbors $x(\mathbf{A}) =
x$, the number of substitutions in a branch of length $t$ is Poissonian with
mean $\mu t x$ and the substitution rate is equal to $\mu x$, as in Kimura's
model. If the variance of the neutral connectivity is not zero, the
substitution distribution is more complicated and has to be computed
numerically using the simulated evolutionary trajectories (see the section
`Materials and Methods').

\section{Results}

An important property that we have shown to hold for the SCN model is that its
statistical properties are \textit{robust}: Their qualitative behavior does
neither depend on the thresholds used to select viable sequences (and does even
not change if lattice models are used instead of effective models of protein
folding with real protein structures), nor it depends on the protein structures
considered, although the studied proteins cover a broad spectrum of different
biological activities. The protein structure only determines which positions
play a structural role and are therefore more conserved than average positions,
but the properties of structurally conserved residues, like the fact that their
evolutionary rate is less dependent on the context of the sequence (see below),
are general features for all structures. Although some properties that we
determined may depend on the length of the protein, our data is insufficient to
quantify such an effect.

We studied in this work eight protein folds, considering for the rubredoxin
fold two different structures, one of a mesophilic and one of a thermophilic
species. The nine proteins studied are: myoglobin (PDB code {\tt 1a6g}),
cytochrome~c (PDB code {\tt 451c}), lysozyme (PDB code {\tt 3lzt}),
ribonuclease (PDB code {\tt 7rsa}), rubredoxin (mesophilic species: PDB code
{\tt 1iro}; thermophilic species: PDB code {\tt 1brf\_A}), ubiquitin (PDB code
{\tt 1u9a\_A}), the TIM barrel (PDB code {\tt 7tim\_A}), and kinesin (PDB code
{\tt 1bg2}). In what follows we describe general properties of the model which
are qualitatively the same for all proteins studied. Unless otherwise stated,
the results that we present refer to the myoglobin fold.

Although the model of protein stability that we use is only an approximate one,
it is noteworthy that our most important qualitative results, the broad
distribution of neutral connectivities and the overdispersion which is caused
by correlations along neutral trajectories (see below), reproduce those of a
former work in which we used a model of protein stability in some sense
complementary to the present one \cite{Bastolla99,Bastolla00b}. In that work
the statistical mechanics of protein sequences was simulated using a lattice
model to represent the ensemble of conformations. Such a treatment provides
only a coarse-grained representation of protein structure (e.g., secondary
structures are not well described). It does, however, allow for the enforcement
of a rigorous criterion of stability, that has been then tested through
extensive Monte Carlo simulations. The fact that qualitatively similar results
are recovered using two complementary approximations to the protein folding
problem makes us confident that our model captures some ``universal''
properties of protein evolution.

\subsection{SCN and structural conservation}

It is commonly observed that protein structures are much more conserved than
amino acid sequences \cite{Holm96,Rost97}, but there is no real evolutionary
justification to impose such a rule as we do in the SCN model. In fact, the
target of natural selection is protein function rather than protein structure,
and the relationship between these two properties is not a simple one. It is
well known that proteins with the same fold can perform quite different
functions, and in some cases there is evidence that those proteins are
evolutionarily related. In fact, proteins often acquire novel functions by gene
duplication though maintaining a very similar structure. Moreover, the case of
structurally unrelated proteins which perform the same function is not rare, a
possible result of convergent evolution.

A more realistic version of the SCN model should impose conservation of protein
thermodynamic stability, irrespective of the native structure. This, in our
opinion, should be a requirement of any model of molecular evolution. We expect
that such an improved version of the SCN model would show that the conservation
of the native structure is a consequence of the stability requirement. The
reason for this expectation is based on our simulations of the present SCN
model and of a previous lattice version of it. In the SCN simulations we record
the smallest value of the normalized energy gap $\alpha$ with the target
conformation, over all neighbors of a sequence belonging to the neutral
network. For seven out of the nine proteins that we studied, the smallest value
of $\alpha$ remained positive for all of the order of $10^5$ examined
sequences. Since a positive $\alpha$ means that the target conformation has the
lowest energy, imposing that the target conformation is stable is effectively
equivalent to imposing that the native (lowest energy) conformation is stable.
Thus stability requirements alone suffice to guarantee the conservation of
these protein structures. The two exceptions that we found are cytochrome~c and
the mesophilic version of rubredoxin, the two smallest proteins that we
studied. For these proteins some sequences in the neutral network have
neighbors where a structure different from the target state has lowest energy.
However, this is not enough to ensure thermodynamic stability of this
structure, since we still have to impose that all structures unrelated to it
have much higher energy, while the energy of the target structure is very low.
We shall investigate this subject in more detail in the future.

From the above considerations it turns out that imposing a well correlated
energy landscape through a condition on the normalized energy gap makes it
difficult to change the native structure maintaining the thermodynamic
stability of the new structure. This result is very different from the one
obtained in the study of neutral networks of RNA. In this case, Schuster and
co-workers have shown through a computational study that the neutral networks
of two different RNA secondary structures can be separated by just one point
mutation \cite{Schuster94}. We think that this difference is not only due to
the difference between protein folding and RNA folding, but also to the
different model of thermodynamic stability applied. Schuster and co-workers
require that the target structure has the lowest energy, while we impose
additionally that it has a large folding free energy and a large normalized
energy gap.

\subsection{Spatial correlations}

\begin{figure}[t]
\centerline{\includegraphics[scale=0.55]{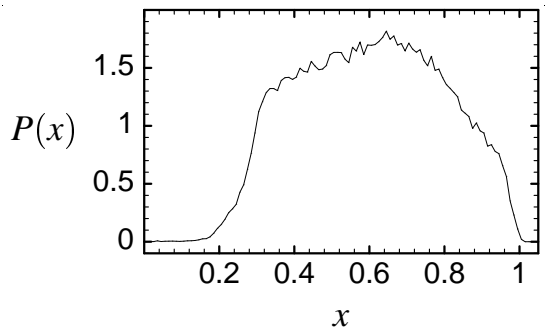}}
\caption{Distribution of the fraction of neutral neighbors for the myoglobin
fold, shown as $P(x)$ vs $x$.}
\label{figure:histogram}
\end{figure}

As we have seen, if the neutral connectivities $x_i$ of different positions are
only weakly correlated, the overall neutral connectivity $x$ has very small
fluctuations and the neutral model of Kimura is valid. However, we observed in
a previous work that the connectivity distribution is quite broad, see
Fig.~\ref{figure:histogram}, which implies that the statistics of the
substitution process is very different from the Poissonian statistics. To
investigate the origin of this important property, we examined the correlation
matrix $\mathcal{C}_{ij}$, whose elements are the correlation coefficients of
the neutral connectivities at positions $i$ and $j$ (see the section `Materials
and Methods').

\begin{figure*}[t]
\centerline{\includegraphics[scale=0.55]{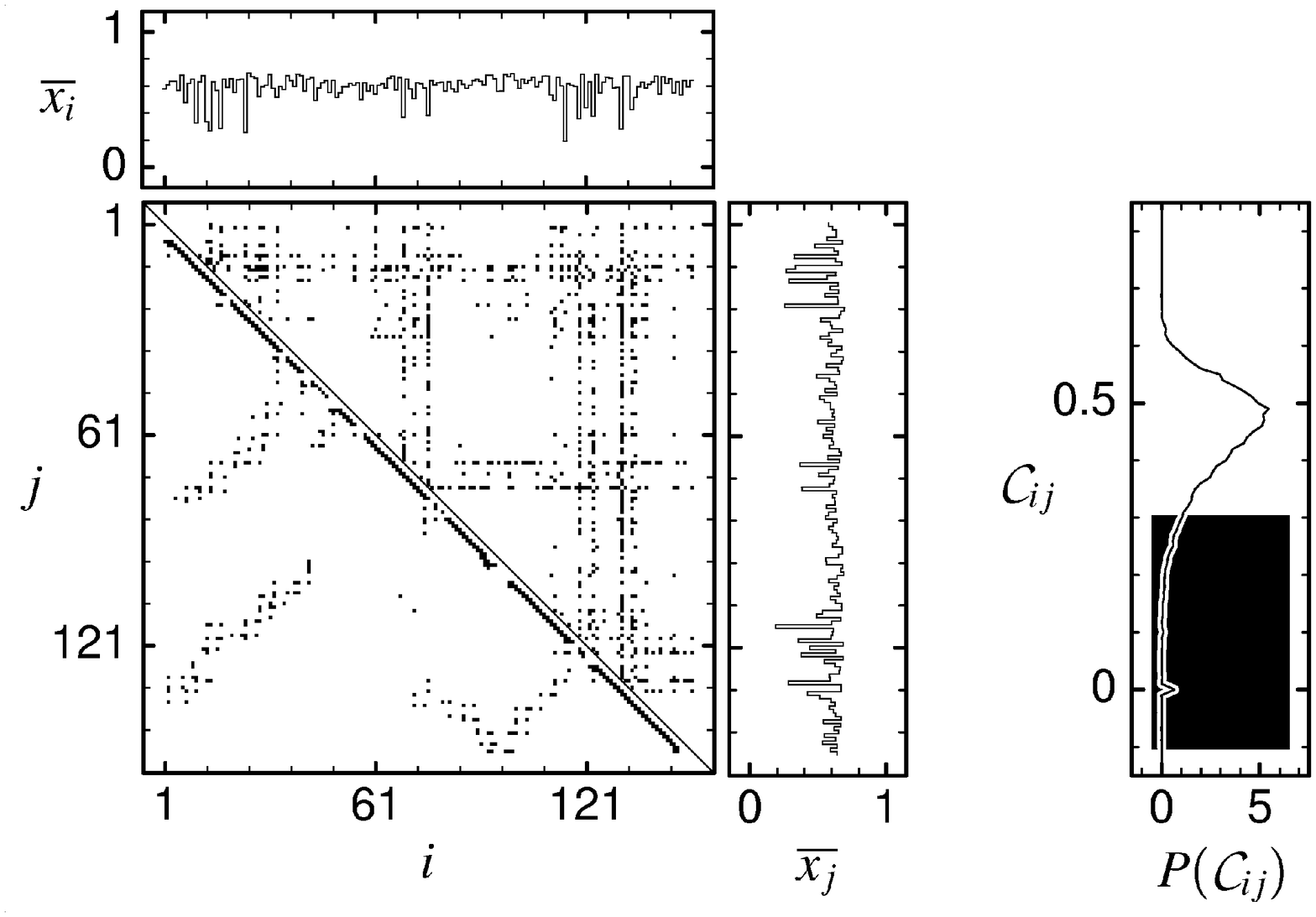}}
\caption{Comparison between cross-correlations and contact matrix. We show the
contact matrix $C_{ij}$ in the lower left part of the central plot (a dot
indicates the residues $i$ and $j$ are in contact) and the cross-correlation
matrix $\mathcal{C}_{ij}$ in the upper right part of the same plot (a black dot
means correlations are weak or absent, $\mathcal{C}_{ij} \le 0.3$, whereas
white means $\mathcal{C}_{ij} > 0.3$). Above and right of the central plot we
show the mean fraction of neutral mutations at position $i$, $\overline{x_i}$.
Conserved positions (low $\overline{x_i}$) show the weakest correlations. The
plot at the very right shows the histogram of the values obtained for the
cross-correlations and the `color code' applied for the central plot.
(Figure has reduced file size and quality, original figure upon request.)}
\label{figure:crosscorrelation+contact}
\end{figure*}

\begin{figure}[t]
\centerline{\includegraphics[scale=0.55]{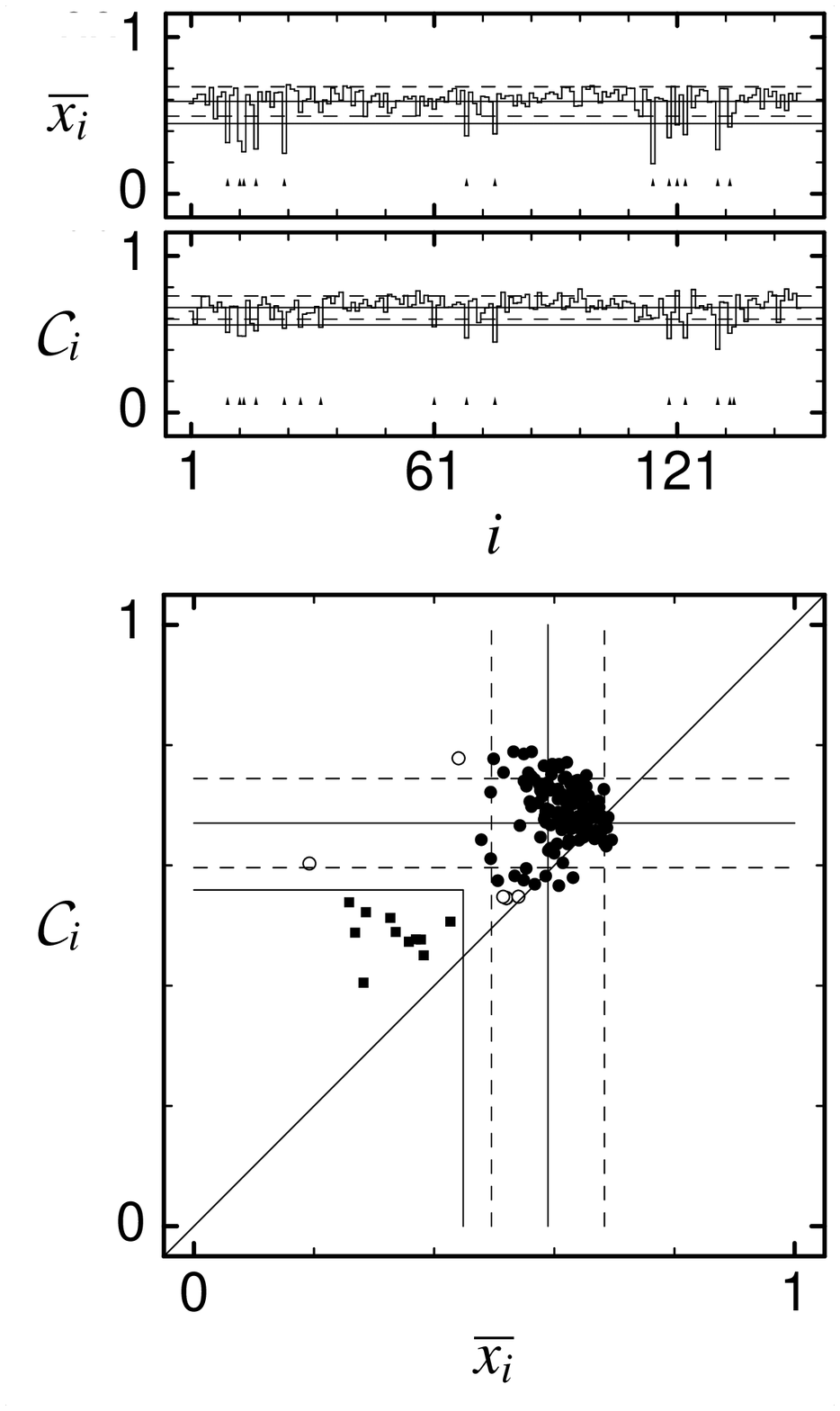}}
\caption{Comparison between cross-correlations and conservation. We show the
mean fraction of neutral mutations at position $i$, $\overline{x_i}$, and the
correlation between $x_i$ and the overall neutral connectivity $x$,
$\mathcal{C}_i$. In the upper part of the figure, the dashed horizontal lines
indicate one standard deviation from the mean (indicated by the full horizontal
lines). The arrows at the bottom indicate the position $i$ for which the mean
fraction of neutral neighbors $\overline{x_i}$ and the cross-correlation
$\mathcal{C}_i$ are below the threshold (1.5 standard deviations) shown as full
horizontal lines. In the lower part of the figure, the two quantities are
plotted against each other (the horizontal and vertical lines have the same
meaning as in the upper plots). The residues which are above the threshold for
both quantities are shown as full circles, the residues which are below the
threshold for both quantities are shown as full squares, whereas the residues
which are above the first threshold but below the second, or vice versa, are
shown as open circles.}
\label{figure:crosscorrelation+conservation}
\end{figure}

Two interesting observations emerge. First, the correlations coefficients
$\mathcal{C}_{ij}$ are positive and large (larger than 0.2) for most positions,
irrespective of whether they are in contact or not in the native structure.
Fig.~\ref{figure:crosscorrelation+contact} shows a comparison between the
correlation matrix $\mathcal{C}_{ij}$ (upper right part of the central plot in
the figure, see the section `Materials and Methods') and the contact matrix of
the native structure $C_{ij}$ (lower left part of the central plot of the
figure). Second, there is a strong relation between correlation and
conservation. Positions which are more conserved, as indicated by the neutral
network average ($\overline{x_i}$) of their fraction of neutral neighbors, show
only a weak correlation with other positions. We plot as black dots in the
upper right part of Fig.~\ref{figure:crosscorrelation+contact} pairs of
positions which are only weakly correlated, $C_{ij} \le 0.3$. These dots
arrange in horizontal and vertical lines at the conserved positions. The
pattern is more clearly shown in
Fig.~\ref{figure:crosscorrelation+conservation}, representing a scatter plot of
the variability index $\overline{x_i}$ versus the correlation coefficient
$\mathcal{C}_i$ between the neutral connectivity at position $i$ and the
overall neutral connectivity (see the section `Materials and Methods'). It can
be seen that the two quantities are strongly correlated.

\citet{Ptitsyn99} showed that there are two groups of conserved residues in the
globin family. In the myoglobin sequence that we considered the first group is
the heme-binding site, formed by the residues Leu29 (B10), Leu32 (B13), Phe33
(B14), Pro37 (C2), Phe43 (CD1), Phe46 (CD4), Leu61 (E4), His64 (E7), Val68
(E11), Leu89 (F4), His93 (F8), Ile99 (FG5), Leu104 (G5), and His142 (H19),
where the number in parenthesis indicates the standard notation of
\citet{Perutz65} that specifies the position within an helix. The second group,
whose conservation is structural and not functional \cite{Ptitsyn99}, is formed
by residues Val10 (A8), Trp14 (A12), Ile111 (G12), Leu115 (G16), Met131 (H8),
and Leu13 (H12). In agreement with our argument, the residues in the second
group are among those with the lowest values of $\mathcal{C}_i$, they are
ranked 9th, 6th, 22th, 24th, 1st, and 14th, respectively. On the contrary,
residues in the functionally conserved group are characterized by average
values of $\mathcal{C}_i$, ranging in rank from from 12 to 132, with an average
rank of 61. Even more interestingly, residue Met131, the one with the lowest
$\mathcal{C}_i$, is anti-correlated (although weakly) with some residues in the
functionally conserved group. This is the only case of significant
anti-correlation. Leu69 and Leu76 are ranked 5th and 2nd, respectively. They
have nearly zero correlation among themselves and with residues in the two
groups. Interestingly, they are located between the two groups in the native
structure. Residues His119 and Phe123 are ranked 3rd and 4th. They are in the
loop between helices G and H and they are probably important to stabilize the
second group of conserved residues. A similar observation holds for residues
Val13 and Ala134, ranked 7th and 8th, respectively, since they are neighbors of
residues Trp14 and Leu115. Taken together this data suggests that the analysis
of cross-correlations complements Ptitsyn's conservation analysis of
structurally important residues \cite{Shakhnovich96,Ptitsyn99}.

\subsection{Temporal correlations}

\begin{figure}[t]
\centerline{\includegraphics[scale=0.55]{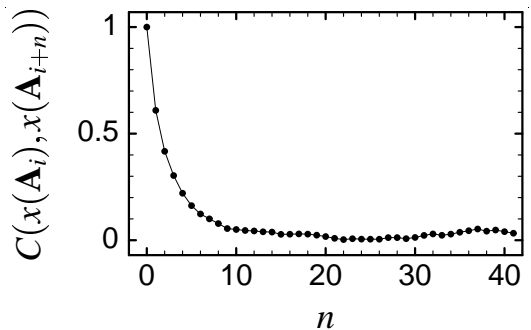}}
\caption{Auto-correlation function $C(x(\mathbf{A}_i),x(\mathbf{A}_{i+n}))$ of
neutral connectivities at sequences separated by $n$ substitutions.}
\label{figure:autocorrelation}
\end{figure}

To investigate the temporal correlations, we measured the auto-correlation
function
\begin{equation}
C(x(\mathbf{A}_i),x(\mathbf{A}_{i+n})) = \frac{1}{m}\sum_{k=1}^m
\frac{x(\mathbf{A}_k) \, x(\mathbf{A}_{k+n}) - \overline{x}^2}{\sigma_x^2}\, ,
\end{equation}
where an average is taken over all pairs of sequences connected through an
evolutionary trajectory of exactly $n$ substitutions. The corresponding plot
can be seen in Fig.~\ref{figure:autocorrelation}. The auto-correlation function
starts with the value one at $t=0$ and then decreases as subsequent
substitutions make the sequences less and less correlated.

The characteristic time after which the correlation function reduces of a
factor $1/\mathrm{e}$ can be estimated through an exponential fit of the first
part of the correlation function. For all proteins examined, the characteristic
length is two or three substitutions. Despite this fast decay, temporal
correlations are responsible of many interesting properties of the model, from
the large deviation of the distribution of the number of substitutions with
respect to a Poissonian one to large variations of the neutral substitution
rates in different evolutionary trajectories (see also \cite{Bastolla02b}).

\subsection{Origin of the correlations}

The phenomenon that originates the spatial and temporal correlations that we
have observed in the SCN model can be described as follows. We impose in our
model that some stability indices are above some predefined thresholds in order
to decide that a sequence is viable. Thus, a sequence where all of the
stability parameters are very high will be much more tolerant to mutations than
an average sequence: In this case, every positions will have a high neutral
connectivity, since most amino acid changes at the position will maintain the
protein viable. The only exceptions will be positions which perform a key
structural role, which, as we have seen, are characterized by a much smaller
neutral connectivity and are much less correlated with the neutral connectivity
of the whole protein. The temporal correlations can be explained exactly in the
same way: Since the thermodynamic properties change smoothly with changes in
the sequences, they will remain very correlated after few changes in the
sequences, so that also the neutral connectivities which derive from these
stability parameters will show strong correlations.

\begin{figure}[t]
\centerline{\includegraphics[scale=0.55]{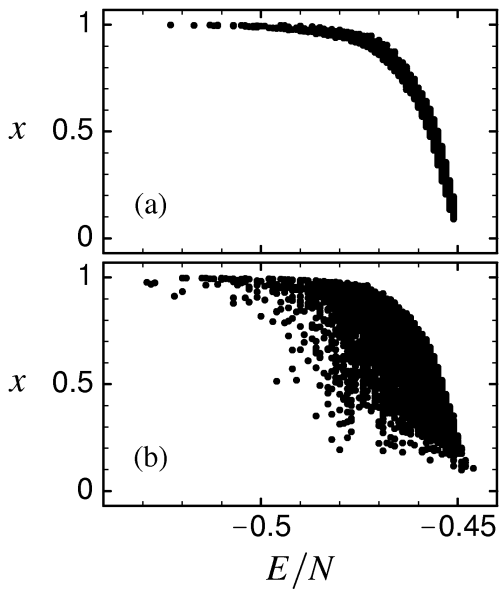}}
\caption{Dependence of the fraction of neutral neighbors $x$ on the native
energy per residue $E/N$ (in arbitrary units), (a)~when only the native energy
per residue has been used as a selection parameter, and (b)~when this parameter
is used together with the normalized energy gap as a selection parameter.
(Figure has reduced file size and quality, original figure upon request.)}
\label{figure:nativeenergy}
\end{figure}

To support our argument, we show results from a simulation where just one
thermodynamic parameter has been considered to select sequences belonging to
the neutral network. In the additional simulations presented in
Fig.~\ref{figure:nativeenergy}(a), only the native energy per residue has been
used as a selection parameter. Accordingly, there is a very strong relationship
between this thermodynamic parameter and the neutral connectivity
$x(\mathbf{A})$. In Fig.~\ref{figure:nativeenergy}(b), both the normalized
energy gap and the native energy have been used as selection parameters. Thus
the relationship between thermodynamic parameters and neutral connectivity is
less strong, but still it is clearly present.

\subsection{Multiple substitutions}

According to the molecular clock hypothesis, the number of substitutions in a
time interval $t$ is proportional to the duration of the time interval, with
the addition of random fluctuations, whose standard deviation is small with
respect to the mean when $t$ is large. However, in the comparison of two
protein sequences, only the number of sites occupied by different amino acids,
$N_{\mathrm{d}}$, is accessible to measurement. From this quantity, the number
of substitutions has to be calculated assuming some model of protein evolution
to correct for possible occurrence of multiple substitutions at the same site.

\begin{figure}[t]
\centerline{\includegraphics[scale=0.55]{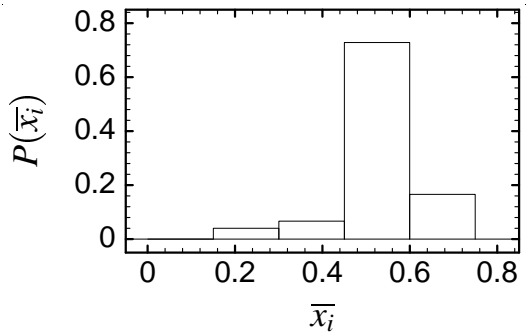}}
\caption{Distribution of the average substitution rate $\overline{x_i}$ across
different positions.}
\label{figure:xi_distribution}
\end{figure}

The simplest such method was used already by \citet{Zuckerkandl62}. Assuming
that all sites in a protein evolve at the same rate, and that the number of
substitutions for each site follows a Poissonian distribution of average value
$K$, the probability that one site did not change is $\exp(-K)$. This
probability can be estimated by the fraction of unchanged sites,
$1-N_{\mathrm{d}}/N$, so that the expected number of substitutions per site $K$
can be estimated as
\begin{equation}\label{eq:PC}
K \approx -\log\left( 1-\frac{N_{\mathrm{d}}}{N} \right)\, .
\end{equation}
The assumption that the substitution rate is the same at all sites is stronger
than the homogeneity assumption stating that the substitution rate is the same
for all sequences. It is well known that functionally important residues change
very rarely during evolution, often as the result of a change in the
environment or a switch in the protein function, while other residues are much
more free to mutate. Even within our model, which does not take into account
functional constraints explicitly, some residues are much more difficult to
mutate than others, since they are structurally important and belong for
instance to the so-called folding nucleus of the protein. In
Fig.~\ref{figure:xi_distribution} we show the distribution of the average
fraction of substitutions at position $i$ which are neutral, for all 151
positions in the myoglobin fold.

As noticed by \citet{Nei00} and by \citet{Kimura83}, the Poissonian correction
Eq.~(\ref{eq:PC}) remains valid also in case of varying evolutionary rates for
small values of the fraction of changed positions, $N_{\mathrm{d}}/N$.
Otherwise, one has to consider more sophisticated corrections. The most common
method which takes into account variation of rate across sites consists in
assuming that the rates are distributed according to a gamma distribution,
which is the product of an exponential times a power law distribution
\cite{Uzzell71}. It is then possible to compute analytically the probability
that a site has not changed and, equating it to the observed frequency, to
obtain the average number of substitutions per site as
\begin{equation}\label{eq:gamma}
K = a \left[ \left( 1-\frac{N_{\mathrm{d}}}{N} \right)^{-1/a}-1 \right] \, ,
\end{equation}
where $a$ is the shape parameter of the gamma distribution \cite{Nei00}. The
above results, Eq.~(\ref{eq:PC}), is recovered in the limit $a\to\infty$, as in
this limit the gamma distribution becomes a delta distribution with vanishing
variance. However, in realistic situations, $a$ tends to be small (typically
smaller than four), so that the distribution of rates across different
positions is broad and the estimated number of substitutions is much larger
than the one obtained using Eq.~(\ref{eq:PC}).

In typical studies, the parameter $a$ is first estimated, for instance by
maximum parsimony, and then used to obtain the number of substitutions through
Eq.~(\ref{eq:gamma}). Our simulations yield the number of substitutions for a
given fraction of mutated positions, $N_{\mathrm{d}}/N$, so that, using our
data, we can directly test the validity of Eq.~(\ref{eq:gamma}).

\begin{figure}[t]
\centerline{\includegraphics[scale=0.55]{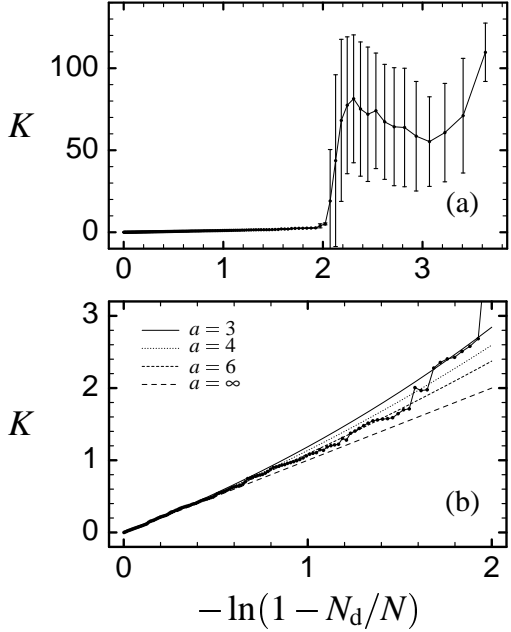}}
\caption{Average number of substitutions $K$ as a function of the logarithm of
the fraction of residues which are the same as in the initial sequence,
$-\ln(1-N_{\mathrm{d}}/N)$. (a)~The transient regime and the stationary regime
are shown. (b)~Only the transient regime is shown, together with the analytical
corrections for multiple substitutions for various values of the shape
parameter $a$, Eq.~(\ref{eq:gamma}). The value $a=\infty$ corresponds to
Poissonian corrections.}
\label{figure:multiplesubstitution}
\end{figure}

Results of such an analysis are shown in
Fig.~\ref{figure:multiplesubstitution}. From
Fig.~\ref{figure:multiplesubstitution}(a) it is clear that two regimes are
encountered. The first regime (large sequence similarity characterized by small
$N_{\mathrm{d}}/N$) will be called the transient regime. In this case the
average sequence similarity decreases on the average with time, measured as the
number of substitutions, and the standard deviation is small. We can thus use
the measured similarity to estimate the number of substitutions. This regime is
illustrated in Fig.~\ref{figure:multiplesubstitution}(b), where it is compared
with various kinds of corrections for multiple substitutions. As it can been
seen, the gamma corrections with small $a$ fits the data better than the
Poissonian correction ($a = \infty$), although systematic deviations are
appreciable. Only for lysozyme we find an optimal value of $a$ close to eight,
whereas for all other proteins the optimal value, although not very precisely
determined, lies in the range from one to five. For low sequence similarities,
it is not allowed to neglect the probability $p\approx 1/20$ that a further
mutation of a mutated site restores the amino acid initially present in the
sequence. Taking this into account, Eq.~(\ref{eq:PC}) has to be modified to
\begin{equation}
K = \ln\left( 1 - p - p K \right) -
 \ln\left( 1-\frac{N_{\mathrm{d}}}{N}-p \right) \, ,
\end{equation}
which, for small $p$, yields
\begin{equation}\label{eq:pi_approx}
K = -(1-p) \ln\left( 1-\frac{N_{\mathrm{d}}}{N (1-p)} \right) \, .
\end{equation}
Similarly, Eq.~(\ref{eq:gamma}) has to be modified to
\begin{equation}
\begin{split}
\left( 1-p-\frac{N_{\mathrm{d}}}{N} \right) =
{} &
\left( 1+\frac{K}{a} \right)^{-a}
\\ & {} \times
\left( 1 - p - p \frac{aK}{a+K}  \right) \, ,
\end{split}
\end{equation}
which for small $p$ can be approximatly solved to yield
\begin{equation}\label{eq:gamma-pi}
K \approx a (1-p)
\left[ \left( 1-\frac{N_{\mathrm{d}}}{N(1-p)} \right)^{-1/a}-1 \right] \, .
\end{equation}
The previous formulas are recovered for $p = 0$, while Eq.~(\ref{eq:pi_approx})
is recovered as the infinite $a$ limit of the above equations. Notice that, for
non-zero $p$, the expected number of substitutions now diverges when the
sequence similarity $1-N_d/N$ tends to $p$, and can not be computed anymore for
smaller sequence similarity.

\begin{figure}[t]
\centerline{\includegraphics[scale=0.55]{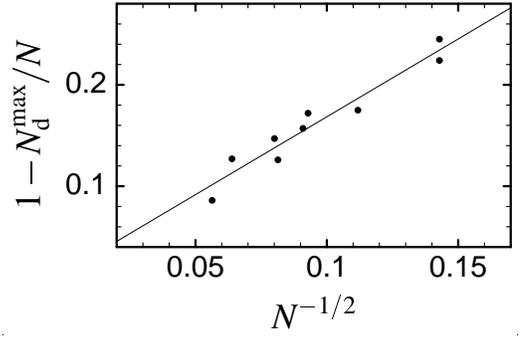}}
\caption{Minimal sequence similarity above which the number of substitutions
can be estimated, $1-N_{\mathrm{d}}^{\mathrm{max}}/N$ vs $N^{-1/2}$, where $N$
is protein sequence length. The data refers to trajectories with $K N=20000$
substitutions. The line shows a linear fit.}
\label{figure:sequencesimilarity}
\end{figure}

The second regime is called the stationary regime (right part of
Fig.~\ref{figure:multiplesubstitution}(a)). In this case, sequence similarity
has reached a stationary distribution with respect to the number of
substitutions, and it does not give us any information about the number of
substitutions occurred in the evolutionary history. Clearly, in this case we
cannot anymore identify the probability that a site has not changed,
$\exp(-K)$, with the frequency of identical sites, $1-N_{\mathrm{d}}/N$,
neglecting the fluctuations of the latter quantity, and consequently the above
formulas loose their validity. The value of sequence similarity below which the
transition to the stationary regime takes place depends on the sequence length
$N$ and, more weakly, on the number of substitutions, $K N$, and is such that
the stationary probability to observe a sequence similarity equal to
$N_{\mathrm{d}}/N$ is roughly equal to $1/K N$. In
Fig.~\ref{figure:sequencesimilarity} we plot the threshold of sequence
similarity below which the standard deviation in the number of substitutions
becomes of the same order of the average number of substitutions, and hence
estimating the number of substitutions looses all meaning. Notice that, if some
positions are conserved because of functional or other constraints, this
threshold of similarity will be even higher. The data has been obtained from
our simulations with $K N = 20000$ substitutions for all nine proteins. The
plot clearly shows the expected dependence of the threshold on the inverse
square root of sequence length.

\subsection{Time dependence of the substitution rate}

\begin{figure}[t]
\centerline{\includegraphics[scale=0.55]{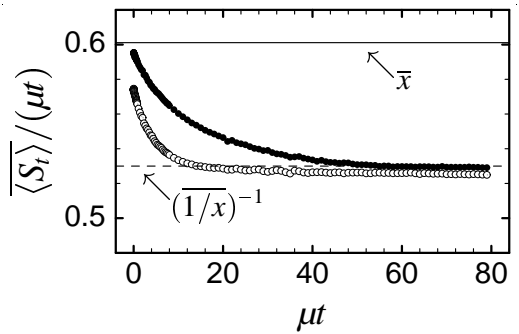}}
\caption{Dependence of the average substitution rate on the time interval
considered. Shown is the average number of substitutions $\overline{\big< S_t
\big>}$ divided by $\mu t$. We present the statistics for the trajectories
obtained through the SCN model (full circles) and the same for the
corresponding annealed trajectories (open circles). The horizontal lines
indicate $\overline{x}$ and $(\overline{1/x})^{-1}$, respectively.}
\label{figure:rate}
\end{figure}

In our analysis, two kinds of averages must be distinguished. We indicate by
angular brackets $\big< \cdot \big>$ the average over the mutation and
acceptance process for a given realization of the evolutionary trajectory, and
by an overline $\overline{\vphantom{S}\cdot\vphantom{S}}$ the average over
evolutionary trajectories. We determine the mean number of substitutions
$\overline{\big< S_t \big>}$ within a time interval $t$, and show this quantity
in Fig.~\ref{figure:rate}. We also show in the same figure results based on
\textit{annealed} trajectories, obtained by extracting at random the values of
$x$ at each substitution event according to the observed distribution $P(x)$
(cf.\ Fig.~\ref{figure:histogram}). In this case, the different $x$ along the
trajectories are independent variables. Note that the annealed trajectories
`interpolate' between the Poissonian case (all $x$ are equal) and the
correlated trajectories obtained through the SCN model: (i)~The comparison
between the annealed trajectories and the simple Poissonian case allows us to
identify the effect of the broad distribution of the number of neutral
neighbors, whereas (ii)~the comparison between the actual and the annealed
trajectories allows us to identify the effect of correlations.

As a result of the broad distribution of connectivities, the substitution rate
$\overline{\big< S_t \big>}/t$ is not constant, as in the standard model of
neutral evolution, but decreases as the time interval $t$ gets longer, as shown
in Fig.~\ref{figure:rate} for both correlated and annealed trajectories. This
can be qualitatively understood by the following consideration: In the annealed
case, the time spans $\tau$ between subsequent substitutions are independent
variables distributed with the density $\overline{D(\tau)} = \int_0^1 P(x) \,
(\mu x)^{-1} \exp(-\mu x \tau) \, \mathrm{d}x$, whose average value is
$\overline{\tau} = \int_0^1 P(x) \, (\mu x)^{-1} \, \mathrm{d}x$. Thus, the
average substitution rate $\overline{\big< S_t \big>}/t$ is not constant in
time as in the Poissonian case. Initially, $S_t$ is a Poissonian variable with
average rate $\mu \overline{x}$. At large time, however, the rate converges to
the smaller value $\overline{\big< S_t \big>}/t \approx 1/\overline{\tau}$
(i.e., $\overline{\big< S_t \big>}/(\mu t) \approx (\overline{1/x})^{-1}$),
since the process spends more and more time in sequences with small
$x(\mathbf{A})$.

\subsection{Variances}

The variance of the substitution process has been intensively studied in the
first tests of the neutral theory. In those tests it was expected that the mean
and the variance of the number of substitutions should be equal under neutral
evolution, as the latter was expected to follow Poissonian statistics. The
ratio between the variance and the mean, $R = \mathrm{V}(S_t)/\mathrm{E}(S_t)$,
where $\mathrm{V}$ indicates variance and $\mathrm{E}$ indicates expectation
value, is called dispersion index. Deviations of this quantity from unity
indicate deviations from Poissonian statistics. In the framework of the SCN
model, the broad distribution of neutral connectivities causes the substitution
process to be overdispersed (its variance is larger than the mean), as we have
shown previously \cite{Bastolla02b}. Here we investigate the relative
contributions of the broad distribution and of the time correlations to the
variance of the substitution process.

The variance of the substitution process can be decomposed in two components,
one calculated for a fixed evolutionary trajectory (fixed population), the
other taking into account the variance of different evolutionary trajectories,
\begin{equation}
\begin{split}
\mathrm{V}(S_t) & = \mathrm{V}_{\mu}(S_t)+\mathrm{V}_x(S_t) \\
& =\left( \overline{\big< S_t^2\big>} - \overline{\big< S_t\big>^2} \right) +
\left( \overline{\big< S_t\big>^2} - \overline{\big< S_t\big>}^2 \right) \, .
\end{split}
\end{equation}
The first term, $\mathrm{V}_{\mu}$, is the variance of the mutation and
acceptance process for a fixed trajectory, averaged over evolutionary
trajectories. The second term, $\mathrm{V}_x$, is the variance of the
substitution rate with respect to different evolutionary trajectories. This
term, which is not present in the standard neutral model, is responsible of the
fact that the variance of the number of substitutions is typically larger than
its mean value, in contrast with a Poissonian process.

We shall denote by $R_{\mu}$ the normalized mutational variance, $R_{\mu}(S_t)
= \mathrm{V}_{\mu}(S_t)/\mathrm{E}(S_t)$, and by $R_x$ the corresponding
normalized trajectory variance, $R_x(S_t) = \mathrm{V}_x(S_t)/\mathrm{E}(S_t)$.
Thus, in the standard neutral model, one expects that $R_{\mu} = 1$ and $R_x =
0$, independent of the time interval $t$.

\begin{figure}[t]
\centerline{\includegraphics[scale=0.55]{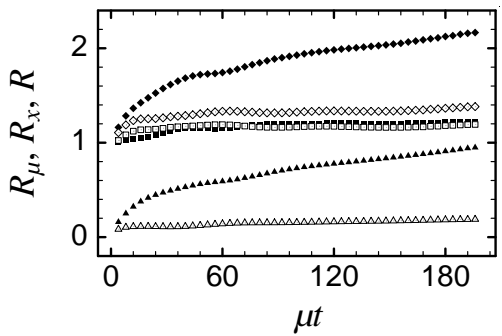}}
\caption{Variances of the substitution process as a function of the time
interval considered. We show the normalized mutation variance $R_{\mu}$
(squares), the normalized trajectory variance $R_x$ (triangles), and the
normalized total variance $R = R_{\mu}+R_x$ (diamonds). We present the
statistics for the trajectories obtained through the SCN model (full symbols)
and the same for the corresponding annealed trajectories (open symbols). Notice
that annealed trajectories, with a broad distribution of neutral connectivities
but lacking temporal correlations, have a total dispersion index only slightly
larger than one, while simulated trajectories have a total dispersion index of
the order of two already after some tens of substitutions.}
\label{figure:variance}
\end{figure}

The results of the substitution process based on simulated trajectories are
shown in Fig.~\ref{figure:variance} as full symbols. We also show as open
symbols results based on annealed trajectories, which have the same
distribution of neutral connectivities but lack the correlations. We start
describing the latter results. As discussed in the previous section, for very
short time interval the substitution distribution is practically Poissonian and
one has $R_{\mu}\approx 1$ and $R_x\approx 0$. As the time interval gets
longer, the two variances grow only moderately: The mutation variance $R_{\mu}$
apparently reaches very soon a stationary value which is just slightly larger
than one for all of the proteins that we studied, while the trajectory variance
$R_x$ has not yet attained a stationary value when the number of mutation
events is of the same order as the protein length, but still remains very
small, of the order of the ratio between the variance and the average value of
the neutral connectivities $x$.

Comparing the two ensembles of trajectories, we note that the presence of
correlations has only a weak effect on the normalized mutation variance
$R_{\mu}$, which remains very close to the value met for annealed trajectories.
However, the normalized trajectory variance $R_x$ increases considerably in
response to the correlations, as $R_x \approx 1$ for $\mu t = 200$.
Surprisingly, $R_x$ even grows with time, although more and more sequences are
used to compute the mutational averages and one could expect that such averages
approach typical values. Thus, surprisingly the large fluctuations between
different trajectories caused by the strong correlations result in a peculiar
phenomenon: Even averaging over an arbitrary long trajectory does not give
values representative of typical trajectories.

The total dispersion index $R=R_x+R_{\mu}$, calculated for the trajectories
obtained through the SCN model, becomes larger than $2$ already for $\mu t
\approx 100$, in qualitative agreement with Gillespie's results for a large set
of proteins, most of which yielded dispersion indices in the range $1 < R
\lesssim 5$ \cite{Gillespie91}. Hence, these large dispersion indices observed
in protein evolution may be to a sizeable extent due to the correlations
present in neutral trajectories in sequence space.

Notice however that, although the dispersion index is larger than one, still
the main property of a Poissonian molecular clock is preserved, since, when the
number of substitutions is large, the average number of substitutions
$\overline{\big< S_t \big>}$ is much larger than the standard deviation
$\sqrt{\mathrm{V}(S_t)}$. Thus fluctuations of the number of neutral
substitutions do not hide completely the pattern arising from the evolutionary
relationships.

\subsection{Lineage effects and the generation time effect}

\begin{figure}[t]
\centerline{\includegraphics[scale=0.55]{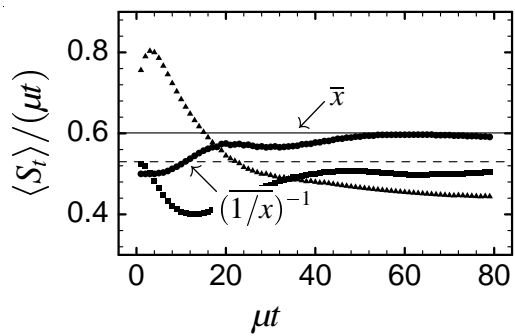}}
\caption{Substitution processes (the average substitution $\big< S_t \big>$)
obtained from three different evolutionary trajectories. The horizontal lines
indicate $\overline{x}$ and $(\overline{1/x})^{-1}$, respectively.}
\label{figure:trajectories}
\end{figure}

The results presented in the previous sections imply the values of $x$ along
one given evolutionary trajectory are very similar for several steps, because
of the strong temporal correlations we identify. Hence, we will observe
trajectories characterized by large substitution rates and trajectories
characterized by small substitution rates, as a result of the broad and
correlated fluctuations in the neutral connectivities. As an example, we show
in Fig.~\ref{figure:trajectories} the substitution rates obtained from three
different evolutionary trajectories. Note that the variance of the neutral
rates $\big< S_t \big>/t$ across different evolutionary trajectories decreases
with the time interval $t$, but rather slowly, so that even after a sizeable
time interval the substitution rates of the three trajectories are still
significantly different.

Since different trajectories can be interpreted as different species, different
substitution rates in different trajectories can simulate lineage effects,
i.e.\ variations of the substitution rate among different taxonomic groups. One
such effect is the generation time effect: Since the natural time unit for
measuring substitution events is the generation, species with longer generation
time are expected to show a slower substitution rate. This prediction has been
verified comparing for instance substitution rates in rodents and primates
\cite{Britten86,Li87,Li00}. Interestingly, the effect is significantly larger
for synonymous substitutions (DNA changes which still code for the same amino
acid) than for non-synonymous ones. Non-synonymous substitutions are
superimposed with the large and correlated fluctuations in the substitution
rate that we just described, and which could be strong enough to obscure the
generation time effect. Thus, the statistics of neutral substitutions observed
here could explain the quantitative difference between synonymous and
non-synonymous substitutions.

\subsection{Tests of neutrality}

A better understanding of the statistical properties of neutral evolution can
help to single out from a background of neutral substitutions the more
interesting cases of positive selection as, for instance, changes in the
protein function and responses to variations of the environment.

The best current bioinformatics method to identify such cases of positive
selection is the one proposed by \citet{McDonald91}, recently used with some
modifications to study the evolution of \textit{Drosophila} genes
\cite{Smith02,Fay02}. This method assumes implicitly that the neutral
substitution process has a constant rate $\mu x$. Thus, the broad and
correlated fluctuations in the neutral connectivities should be taken into
account to provide a better null hypothesis to compare with.

McDonald and Kreitman's method examines two samples of sequences from two
related populations and distinguishes two classes of nucleotide differences in
protein coding genes: intra-population differences (polymorphisms) and
positions which are fixed in any two populations but differ between the two
(fixed differences). Each class is further divided into synonymous (S) and
non-synonymous (A) differences. Let us denote by $D_{\mathrm{PS}}$ and
$D_{\mathrm{PA}}$ the number of synonymous and non-synonymous polymorphisms,
respectively, and by $D_{\mathrm{FS}}$ and $D_{\mathrm{FA}}$ the number of
synonymous and non-synonymous fixed differences, respectively. Under the
neutral hypothesis, polymorphisms and substitutions are just the two faces of
the same coin. Let us suppose that all synonymous mutations are neutral, and
that all deleterious mutations are quickly removed from the population and do
not show as polymorphisms. Under this hypothesis, the ratio
$D_{\mathrm{PA}}/D_{\mathrm{PS}}$ of non-synonymous to synonymous polymorphisms
can be interpreted as the fraction of neutral mutations, $x$, in the two
populations. Analogously, $D_{\mathrm{FA}}/D_{\mathrm{FS}}$ can be interpreted
to represent the fraction of neutral mutations, $x$, over an evolutionary
trajectory. McDonald and Kreitman assume that the two quantities are equal
under neutral evolution. Thus, in their method, the neutral expectation is that
$D_{\mathrm{PA}}/D_{\mathrm{PS}}$ and $D_{\mathrm{FA}}/D_{\mathrm{FS}}$ are
equal. Conversely, if the latter ratio is significantly larger than the former,
this is interpreted as an evidence that some of the non-synonymous
substitutions have been fixed by positive selection, since in this case
fixation is much faster than for neutral mutations, and a larger number of
substitutions can be fixed if they bring any selective advantage.

Let us now see how the neutral expectation must be modified taking into account
the statistics of neutral substitutions. The ratio of non-synonymous to
synonymous polymorphisms in two populations can be interpreted as the fraction
of neutral mutations in the two populations, $x(\mathbf{A}_1)$ and
$x(\mathbf{A}_2)$ respectively. On the other hand, the ratio
$D_{\mathrm{FA}}/D_{\mathrm{FS}}$ should be interpreted as the average number
of substitutions taking place in an evolutionary trajectory whose length $t$ is
double of the time spent since the last common ancestor of the two
populations,
\begin{equation}\label{eq:D_F}
\frac{D_{\mathrm{FA}}}{D_{\mathrm{FS}}} \approx
\frac{\big< S_t \big>}{\mu t} \, ,
\end{equation}
where the average is the mutational average along the evolutionary trajectory
which connects the two wild-type sequences $\mathbf{A}_1$ and $\mathbf{A}_2$.
Now, if the number of substitutions between these two sequences is small with
respect to the ``correlation time'' after which values of $x(\mathbf{A})$
become uncorrelated, then we expect that the effective fraction of neutral
mutations represented in Eq.~(\ref{eq:D_F}) is close to both $x(\mathbf{A}_1)$
and $x(\mathbf{A}_2)$, and the neutral expectation of McDonald and Kreitman
remains valid. Otherwise, we expect that $D_{\mathrm{FA}}/D_{\mathrm{FS}}$ is
smaller than the larger between the two neutral fractions: As shown in
Fig.~\ref{figure:rate}, the effective substitution rate decreases as longer
time intervals are considered, because the evolutionary process gets trapped in
sequences with small neutral connectivity. As a rule of thumb, if
$x(\mathbf{A}_1)$ and $x(\mathbf{A}_2)$ are very similar and the number of
fixed substitutions $D_{\mathrm{FA}}$ is of the order of 10, then the neutral
expectation of the McDonald and Kreitman method is likely to be valid,
otherwise corrections should be considered.

\section{Conclusions and perspectives}

In this work we have proven that the statistical properties of neutral
substitutions simulated through the SCN model, which imposes conservation of
the thermodynamic stability of the protein, differ considerably from those
predicted by the standard Kimura's model, which assumes that the fraction of
neutral mutations is constant throughout evolution. These differences have
important consequences for several aspects of neutral evolution. Only taking
into account accurate statistical properties of a neutral model allows to test
the neutral hypothesis as the null hypothesis of molecular evolution, against
which adaptive substitutions have to be distinguished.

We have shown that the differences between the standard neutral model and the
SCN model arise because of correlations: (i)~There are strong spatial
correlations between the neutral connectivities of different positions along
the protein chain, even if the corresponding amino acids are not close in
space. As a consequence, the neutral connectivities of the entire sequences are
broadly distributed, in contrast with the standard model. (ii)~There are strong
temporal correlations between the neutral connectivities of sequences nearby
along a neutral trajectory. As a consequence, the substitution rates of
different evolutionary trajectories, even averaged over a sizeable number of
substitutions, can be significantly different.

We have illustrated a number of important consequences of these properties:
(i)~As a result of the broad distribution, the substitution rate is a
decreasing function of the time interval considered, instead of being constant,
and attains the stationary value $\mu (\overline{1/x})^{-1}$ starting from the
larger value $\mu \overline{x}$ at very small time intervals. (ii)~As a result
of the auto-correlations, even long evolutionary trajectories, corresponding to
different species, can be characterized by rather different neutral
substitutions rates. This fact can produce new lineage effects which may
obscure the generation time effect, and it should be taken into account when
testing the predictions of the neutral hypothesis. (iii)~As a result of the
broad variance of different evolutionary trajectories, the variance of the
substitution process is not constant in time, and it is larger than that
expected for a Poissonian process. Nevertheless, for large evolutionary
separations the standard deviation of the number of substitutions is much
smaller than the mean value, as for a Poissonian process, so that neutral
substitutions can still be useful for reconstructing phylogenetic trees.
(iv)~Moreover, our simulations provide useful data to test methods for
estimating the number of substitutions between two diverging sequences from
their sequence similarity. Our results allow to estimate the threshold below
which sequence similarity does not provide any information on the number of
substitutions, and also show that the probability that the same amino acid
arises as the result of two independent substitutions should be taken into
account at low similarity. The neutral model does not make detailed predictions
about this point, since one still needs to know the neutral mutation rates
$x_i$ at different positions. It is important to note that Kimura's model does
not assume that such rates are all equal. Nevertheless, our simulations produce
realistic $K$ vs $N_{\mathrm{d}}/N$ curves which can be used to test correction
methods.

The structural approach to protein evolution has only just been initiated.
Natural extensions of our model that will be considered in forthcoming work
include consideration of the genetic code as well as modeling of the population
genetics level and of possible fitness effects of changes in the thermodynamic
parameters of the protein. We anticipate that the structural approach will
provide key insights into different areas of molecular sciences, including
directed evolution of enzymes, rational sequence design, phylogenetic
reconstruction, protein fold prediction, and production of new materials for
bio-nanotechnology.

\section{Materials and Methods}

\subsection{Protein model}

We represent a protein structure by its contact matrix $C_{ij}$, where $C_{ij}
= 1$ if residues $i$ and $j$ are in contact and $C_{ij} = 0$ otherwise. Two
residues are considered in contact if any two of their heavy atoms are closer
than $4.5 \, \textrm{\AA}$. The effective free energy associated to a sequence
of amino acids $\mathbf{A}$ in the configuration $\mathbf{C}$ is approximated
as a sum of pairwise contact interactions,
\begin{equation}
E(\mathbf{A},\mathbf{C}) = \sum_{i<j} C_{ij} U(A_i,A_j) \, ,
\end{equation}
where $A_i$ labels one of the twenty amino acid types and $U(\cdot,\cdot)$ is a
$20\times 20$ symmetric interaction matrix. Here we use the matrix derived by
\citet{Bastolla00b}, which describes accurately the thermodynamic stability of
a large set of monomeric proteins \cite{Bastolla01}.

Three remarks are needed: First, the effective energy parameters implicitly
take into account the effect of the solvent and depend on temperature. They
express free energies rather than energies. Second, the effective energy of a
structure is defined with respect to a completely extended reference structure
where no contacts are formed and which sets the zero of the energy scale.
Third, one can derive from the database not the parameters $U$ themselves but
the dimensionless quantities $U/(k_{\mathrm{B}} T)$. It is thus important to
use dimensionless parameters to evaluate the stability of the protein model.

\subsection{Candidate structures}

We generate candidate structures for a protein sequence of $N$ residues by
generating all possible gapless alignments of the sequence with structures in
the PDB. This procedure is called \textit{threading}. In this way, we generate
many, typically of the order of $10^6$, protein-like structures per sequence.
In the present context, threading is directly used to produce the contact maps
of the candidate structures. In order to speed up the computations, we use a
non-redundant subset of the PDB excluding proteins with homologous sequences,
selected by \citet{Hobohm94}.

\subsection{The folding parameter $\alpha$}

For a given sequence $\mathbf{A}$, the energy landscape is well correlated if
all configurations of low energy are very similar to the configuration of
minimal effective energy, $C^*(\mathbf{A})$. Structure similarity is measured
by the overlap $q(\mathbf{C},\mathbf{C}^*)$, counting the number of contacts
that two structures have in common and normalizing it through the maximal
number of contacts, so that $q$ is comprised between zero and one. In a well
correlated energy landscape, the inequality
\begin{equation}\label{eq:alpha}
\frac{E(\mathbf{A},\mathbf{C})-E(\mathbf{A},\mathbf{C}^*)}
{|E(\mathbf{A},\mathbf{C}^*)|} \geq \alpha(\mathbf{A})
\left( 1-q(\mathbf{C},\mathbf{C}^*) \right)
\end{equation}
holds, stating that the energy gap between each alternative structure
$\mathbf{C}$ and the ground state $\mathbf{C}^*$, measured in units of the
ground state energy, is larger than a quantity $\alpha(\mathbf{A})$ times the
structural distance $1-q(\mathbf{C},\mathbf{C}^*)$. The dimensionless quantity
$\alpha(\mathbf{A})$, which is the largest quantity for which the above
inequality holds, can be used to evaluate the folding properties of sequence
$\mathbf{A}$. For random sequences, many different configurations have quite
similar energy and hence $\alpha(\mathbf{A})\approx 0$. In this case the energy
landscape is rugged, the folding kinetics is very slow, and the thermodynamic
stability with respect to variations in the solvent is very low. In contrast,
computer simulations of well designed sequences have shown that, when
$\alpha(\mathbf{A})$ is finite, the folding kinetics is fast and the stability
with respect to changes in the energy parameters as well as mutations in the
sequence is very high.

Our algorithm computes the parameter $\alpha(\mathbf{A})$ for a fixed target
configuration $\mathbf{C}^*$ and a large number of sequences $\mathbf{A}$. We
thus indicate this parameter as $\alpha(\mathbf{A},\mathbf{C}^*)$, since we do
not know a priori that $\mathbf{C}^*$ has lowest energy. Notice however that,
if $\alpha(\mathbf{A},\mathbf{C}^*)$ is positive, all alternative structures
have higher energy than $\mathbf{C}^*$. We impose that
$\alpha(\mathbf{A},\mathbf{C}^*)$ is larger than a positive threshold
$\alpha_{\mathrm{thr}}$ for sequences $\mathbf{A}$ belonging to the neutral
network.

\subsection{The $Z$-score}

The $Z$-score $Z(\mathbf{A},\mathbf{C}^*)$ \cite{Bowie91,Goldstein92} is a
measure of the compatibility between a sequence $\mathbf{A}$ and a structure
$\mathbf{C}^*$, widely used in structure prediction. It depends on an effective
energy function, and measures the difference between the energy of sequence
$\mathbf{A}$ in configuration $\mathbf{C}^*$ and its average energy in a set of
alternative configurations, $\{ \mathbf{C}\}$, in units of the standard
deviation of the energy,
\begin{equation}
Z(\mathbf{A},\mathbf{C}^*) =
\frac{E(\mathbf{A},\mathbf{C}^*)-
 \big< E(\mathbf{A},\mathbf{C})\big>_{\mathbf{C}}}
{\sqrt{\big< E(\mathbf{A},\mathbf{C})^2\big>_{\mathbf{C}}-
\big< E(\mathbf{A},\mathbf{C})\big>_{\mathbf{C}}^2}} \, .
\end{equation}
When sequence $\mathbf{A}$ folds in structure $\mathbf{C}^*$, its corresponding
$Z$-score is very negative.

Given the above definition, one has still to specify how to select alternative
structures. A possibility, often used for lattice models \cite{Mirny96} is to
assume that alternative structures are maximally compact, randomly chosen
structures, whose average energy can be estimated as $\big<
E(\mathbf{A},\mathbf{C}) \big>_{\mathbf{C}} = N c_{\mathrm{max}} \big<
e(\mathbf{A}) \big>$. Here, $N c_{\mathrm{max}}$ is the maximal number of
contacts of candidate structures and $\big< e(\mathbf{A}) \big>$ is the average
energy of a contact, averaged over all possible contacts formed by sequence
$\mathbf{A}$. This leads to introduce the parameter
\begin{equation}
Z'(\mathbf{A},\mathbf{C}^*) =
\frac{E(\mathbf{A},\mathbf{C}^*)/
 N c_{\mathrm{max}}-\big< e(\mathbf{A})\big>}
{\sqrt{\big< e^2(\mathbf{A})\big>-\big< e(\mathbf{A})\big>^2}} \, .
\end{equation}
The use of $Z'$ has two main advantages: (i)~It makes the value of the
$Z$-score much less sensitive to chain length $N$ and to the particular set of
alternative structures used. (ii)~The evaluation of $Z'$ is much faster than
that of the $Z$-score. This is necessary in order to explore efficiently
sequence space. We impose that $-Z'(\mathbf{A},\mathbf{C}^*)$ is larger than a
positive threshold $-Z_{\mathrm{thr}}$ for sequences $\mathbf{A}$ belonging to
the neutral network.

\subsection{Sampling the neutral network}

Our algorithm explores the neutral network of a given protein starting from its
PDB sequence $\mathbf{A}_1$ and iterating the following procedure: At iteration
$i$, (i)~the number $X(\mathbf{A}_i)$ of viable neighbors of sequence
$\mathbf{A}_i$ is computed, and (ii)~the sequence $\mathbf{A}_{i+1}$ is
extracted at random among all the viable neighbors of $\mathbf{A}_i$. In this
way we generate a stochastic process along the neutral network which simulates
neutral evolution and looses memory of the initial sequence very fast.

Sequence $\mathbf{A}$ is regarded as viable if both parameters
$\alpha(\mathbf{A},\mathbf{C}^*)$ and $-Z'(\mathbf{A},\mathbf{C}^*)$ are above
predetermined thresholds, chosen as 98.5\% of the values of those parameters
for the sequence in the PDB. This enforces conservation of the thermodynamic
stability and folding capability of the native structure $\mathbf{C}^*$. We
verified that the observed behavior does not change qualitatively for
thresholds between 95\% and about 100\% of the PDB values.

We impose strict conservation of the cysteine residues in the PDB sequence, and
do not allow any residue to mutate to cysteine, since a mutation changing the
number of cysteine residues by one would leave the protein with a very reactive
unpaired cysteine that would probably affect its functionality. Accordingly,
the maximal possible number of neighbors tested is $X_{\mathrm{max}} = 18
(N-N_{\mathrm{cys}})$, where $N$ is the number of residues and
$N_{\mathrm{cys}}$ is the number of cysteine residues in the starting sequence.
The total number of viable point mutations, $X(\mathbf{A})$, expresses the
local connectivity of the neutral network. We normalize it by the total number
of neighbors, $X_{\mathrm{max}}$, getting the fraction of neutral neighbors,
$x(\mathbf{A}) = X(\mathbf{A})/X_{\mathrm{max}}\in (0,1]$.

To compute $x(\mathbf{A})$, we have to evaluate the $\alpha$ parameter for all
sequences $\mathbf{A}'$ obtained through a point mutation of sequence
$\mathbf{A}$. From Eq.~(\ref{eq:alpha}) we note that the $\alpha$ parameter can
be obtained from the configuration with the highest destabilizing power, i.e.\
the highest value of the energy gap divided by the structural distance from the
native configuration. These change from sequence to sequence, but it is
expected not to change very much for neighboring sequences. Thus, in order to
speed up the computation of $\alpha(\mathbf{A}',\mathbf{C}^*)$, instead of
considering all candidate configurations we consider only the 50 configurations
with the highest destabilizing power (i.e.\ the energy gap divided by the
structural distance from the native configuration) for sequence $\mathbf{A}$
and compute their mutated destabilizing power using sequence $\mathbf{A}'$. The
$\alpha$ parameter is then obtained from the configuration with the highest
destabilizing power. This procedure could slightly overestimate
$\alpha(\mathbf{A}',\mathbf{C}^*)$ since not all configurations are used, but
we have checked that the error introduced in the $x$ value is in all cases
below 0.1\%.

\subsection{Cross-correlations}

To obtain a deeper insight into the mechanism of neutral evolution, it is
helpful to study the cross-correlations between the fraction of neutral
neighbors for two given residues of a protein of $N$ residues. The analysis
starts from $N$ individual trajectories $\{ x_i(\mathbf{A}_1),
x_i(\mathbf{A}_2), \ldots \}$ of $m$ evolutionary steps for each residue $i$.
We define the average fraction of neutral neighbors $\overline{x_i} = (1/m)
\sum_{k=1}^m x_i(\mathbf{A}_k)$ and the corresponding variance $\sigma_i^2 =
\overline{x_i^2} - \overline{x_i}^2$ for each residue $i$. Then, we calculate
the correlation matrix $C_{ij}$, where $\mathcal{C}_{ij} = \mathcal{C}_{ji}$
determines the cross-correlations between residue $i$ and $j$ and is defined as
\begin{equation}
\mathcal{C}_{ij} = \frac{1}{m} \sum_{k=1}^m
\frac{\Big( x_i(\mathbf{A}_k)-\overline{x_i} \Big)
\Big( x_j(\mathbf{A}_k)-\overline{x_j} \Big)}
{\sigma_i \sigma_j} \, .
\end{equation}
Due to the enforced conservation of cysteine, these residues require a special
treatment: If residue $i$ is cysteine, then $\mathcal{C}_{ij} =
\mathcal{C}_{ji} = 0$ for all $j$.

One needs to distinguish three different cases, (i)~$\mathcal{C}_{ij} = 0$
means that the fraction of neutral neighbors for residues $i$ and $j$ are
uncorrelated, (ii)~$\mathcal{C}_{ij} > 0$ indicates that the fraction of
neutral neighbors for residues $i$ and $j$ are correlated, and
(iii)~$\mathcal{C}_{ij} < 0$ shows that the fraction of neutral neighbors for
residues $i$ and $j$ are anti-correlated.

Alternatively, one may also study the cross-correlations between the fraction
of neutral neighbors for one given residue and the fraction of neutral
neighbors for the whole protein. Defining the average fraction of neutral
neighbors for the whole protein $\overline{x} = (1/m) \sum_{k=1}^m
x(\mathbf{A}_k)$ and the corresponding variance $\sigma^2 = \overline{x^2} -
\overline{x}^2$, we calculate the correlation vector $\mathcal{C}_i$ which
determines the cross-correlations between the fraction of neutral neighbors for
residue $i$ and the fraction of neutral neighbors for the whole protein,
\begin{equation}
\mathcal{C}_i = \frac{1}{m} \sum_{k=1}^m
\frac{\Big( x_i(\mathbf{A}_k)-\overline{x_i} \Big)
\Big( x(\mathbf{A}_k)-\overline{x} \Big)}{\sigma_i \sigma} \, .
\end{equation}
Here, again three different cases need to be distinguished, (i)~$\mathcal{C}_i
= 0$ indicates that the fractions of neutral neighbors for residues $i$ and for
the whole protein are uncorrelated, (ii)~$\mathcal{C}_i > 0$ means that the
fractions of neutral neighbors for residues $i$ and for the whole protein are
correlated, and (iii)~$\mathcal{C}_i < 0$ shows that the fractions of neutral
neighbors for residues $i$ and for the whole protein are anti-correlated.

\subsection{Substitution process}

Given an evolutionary trajectory $\{ x(\mathbf{A}_1), x(\mathbf{A}_2), \ldots
\}$, the distribution of the number of substitutions taking place in a time $t$
can be computed by considering Eq.~(\ref{eq:Pacc}), where $k$, the number of
attempted mutations, is a Poissonian variable of average value $\mu t$.

In order to handle the computation, we divide all values of $x$ in $M$ classes,
choosing $x_c$ as representative value of all $x$'s belonging to class $c$. The
number of operations needed to evaluate the substitution probability increases
exponentially with the number of classes $M$. At the same time, the evaluation
becomes more and more accurate as $M$ increases. We chose $M=6$ in our
numerical computations as a reasonable compromise between accuracy and
rapidity, checking that larger values of $M$ introduce only small changes.

\onecolumngrid\vspace*{10mm}
\noindent\textbf{Abbreviations:}
PDB: Protein Data Bank,
SNC model: Structurally Constrained Neutral Model
\twocolumngrid

\end{document}